\documentstyle[12pt]{article}
\def\d{{\rm d}}
\begin{document}
\begin{titlepage}
\begin{flushright}
TIFR/TH/94-41,\\
30 Sept. 94.
\end{flushright}
\begin{center}
\vspace{0.4cm}
\Large {\bf Photon Structure Functions: Target Photon Mass Effects and
QCD Corrections}
\vspace{1.0cm}
\normalsize

\centerline{\large\bf Prakash Mathews$^{\dag}$}
\centerline{\em Centre for Theoretical Studies}
\vspace{-.2cm}
\centerline{\em Indian Institute of Science}
\vspace{-.2cm}
\centerline{\em Bangalore 560 012, INDIA.}
\vspace{.2cm}
\centerline {and}
\vspace{.2cm}
\centerline{\large\bf V. Ravindran$^{\ddag}$}
\centerline{\em Tata Institute of Fundamental Research}
\vspace{-.2cm}
\centerline{\em Homi Bhaba Road, Bombay 400 005, INDIA.}
\vspace{1.0cm}

\centerline{\bf Abstract}
\begin{quote}
\small{
We present a systematic analysis of the processes $e^+ ~e^-
\rightarrow e^+ ~e^- X$ to study the polarised and unpolarised
photon structure functions.  The effect of target photon mass,
which manifests itself as new singly polarised structure functions is
studied.  The physical interpretation of these structure functions
in terms of hadronic components is given using the free field
analysis.  Assuming factorisation of the photon structure tensor,
the relevant QCD corrections to the various structure functions
are evaluated.  The effect of the target photon mass and QCD
corrections on the unpolarised and polarised cross sections are
also studied}
\end{quote}
\end{center}

\vspace{1cm}
\begin{flushleft}
e-mail:\\
\nopagebreak
$\dag$ prakash@cts.iisc.ernet.in\\
$\ddag$ ravi@theory.tifr.res.in
\end{flushleft}
\end{titlepage}

\section{Introduction}
The collision of photons at high energy electron-positron
colliders is yet another comprehensive laboratory for
testing Quantum Chromodynamics (QCD).  The process $ e^+
{}~e^- \rightarrow e^+ ~e^- X$ (hadrons) (Fig.~1), at very
high energies can be studied in terms of the hadronic structure
of the photon.  This process is related to the photon-photon
$\rightarrow $ hadrons subprocess.  In the past theoretical
aspects of this process have been studied by various authors\cite {PAN}.
Ahmed and Ross \cite {AR} had studied the photon structure
function by considering the $\gamma ~\gamma$ point scattering
and used the Operator Production Expansion (OPE) to take the
non perturbative effects into account.  The behaviour of this
process in the context of perturbative QCD was first studied
using the OPE by Witten {\cite{EW}}, who showed that the
unpolarised structure function $F_2^\gamma$ increases as $\ln
Q^2$, where $Q^2\equiv -q^2$ and $q$ is the momentum of the
probing photon.  A complete next to leading order QCD corrections
to unpolarised structure function can be found in \cite{EL}.  The
phenomenological implications of this process in terms of single
and double resolved photons were highlighted by Drees and Godbole
\cite{DG}.  In the recent past the polarised structure function
has attracted a lot of attention \cite{AM,ET,SDB,NSV}.  In \cite
{AM} the OPE analysis has been extented to the polarised sector,
while in \cite{ET,SDB} the first moment of polarised structure
function has been evaluated and found to be zero for a massless
photons.  The sensitivity of this sum rule due to the off shell
nature of the target photon has been addressed in \cite{NSV}.

Experimentally the process $e^+ e^- \rightarrow e^+ e^- X$ has been
studied for over a decade, but much of the present interest is due
to the recent experiments at LEP and TRISTAN.  Both LEP \cite{LEP}
and TRISTAN \cite{AMY} have recently reported three or more jets
in the final state of $\gamma~ \gamma \rightarrow {\rm hadrons}$.
This indicates the contribution from additional QCD subprocesses.  LEP
\cite{LEP1} has measured the hadronic photon structure function
$F^\gamma_2 $ and has observed significant point like components
of the photon when the probe photon has $Q^2 > 4$ GeV$^2$.  TRISTAN
\cite{AMY1} has measured the inclusive cross section of jets in the
$\gamma~ \gamma$ interactions and reports that the jet rates show
evidence for hard scattering effect of hadronic constituents of photon.
The differential cross section of $e^+ e^-\rightarrow e^+ e^- X$
process has a $1/k^2$ pole due to the soft photon, where $k$ is the
momentum of the target photon.  So the experiment will be sensitive
to the off shell nature of the target photon.  {\it The off shell
nature of the photon gives rise to zero polarisation which in turn
induces additional structure functions to fully characterise the
photon}.  Hence a comprehensive study of the photon structure
functions at this stage including the QCD and the mass effects
of the photons is much awaited.
Future experimental prospects are also bright at the Next Linear
Colliders (NLC) where high energy photon beams can be produced by back
scattering a high intensity laser beam on a high energy $e^-$ beam
\cite{IG}.  In this process each electron in the bunch can Compton
scatter to yield a nearly collinear beam of photons with a high
fraction of $e^-$ beam energy.  The advantage of the back scattered
laser beam is that the photon luminosity is comparable to that of
$e^+ e^-$, unlike the beamstrahlung and equivalent photon processes.

In the process we have considered, the subprocess photon-photon
$\rightarrow$ hadrons, a large off-shell beam photon probes a
target photon which is off shell.  We {\it treat} the target
photon as a composite object consisting of both hadronic and
photonic components.   In our analysis, we have treated the
photon all most on par with the proton.  This would imply that
the probe photon would in fact see quarks and gluons (hadronic
components) as well as photons in the off shell target photon
which can be produced at higher orders.  All the higher order
effects that go into the productions of quarks, gluons and
photons in the target photon are treated as a blob as shown in
the Fig.~5.  Hence we are now in a position to utilise all
the machinery that is used in the lepton-proton Deep Inelastic
Scattering (DIS).  Due to the off shell nature of the target
photon, additional polarisation of the target gives rise to
new structure functions to completely characterise the process.
These new structure functions are the singly polarised structure
functions.  In the context of the photon, these new structure
functions arising due to zero polarisation of the target photon
(mass effect) are being considered for the first time.  To
understand the structure functions in terms of the hadronic
components we perform a free field analysis of this process.
In doing so we arrive at various sum rules and relations among
the various photon structure functions.  These structure
functions are related to the photonic matrix elements of some
bilocal operators similar to those one comes across in DIS and
Drell-Yan.  We assume the factorisation of hard and soft parts
in order to calculate the hadronic and photonic contributions
to the cross section.  Our approach is different from the
previous analysis which usually uses the parton model picture
or OPE.

\section{Target photon mass effects}
Consider the process $e^- (p_1,s_1)~ e^+ (p_2,s_2) \rightarrow e^-
(p_1^\prime) ~e^+ (p_2^\prime)~ X(p_{\bf x})$, where $s_1$, $s_2$ are
the polarisation vectors of the leptons  and $X$ represents the final
state hadrons.  Let us work in the $cm$ frame of the incoming leptons.
The momenta of the in coming and out going particles are parametrised
as $p_1=(E_1,0,0,E_1)$,
$p_2=(E_1,0,0,-E_1)$ and $p_1^\prime = E_1^\prime (1,0,\sin \theta_1,
\cos \theta_1)$, $p_2^\prime =E_2^\prime(1,0,-\sin \theta_2,-\cos
\theta_2)$ respectively.  $\theta_{1,2}$ are the scattering angles of out
going leptons with respect to the beam axis. The total cross section
for this process is given by
\begin{eqnarray}
\d \sigma=\frac{1}{8 E_1^2}~\frac{\d^3p_1^\prime}{(2\pi)^3 2 E_1^\prime}
{}~\frac{\d^3p_2^\prime}{(2\pi)^3 2 E_2^\prime}~\prod_{i=1}^n
{}~\frac{\d^3p_i}{(2\pi)^3 2 E_i} ~\vert T_n \vert^2 ~(2 \pi)^4 \delta^4(
p_1+p_2-p_1^\prime-p_2^\prime-p_n)~,
\end{eqnarray}
where $T_n$ is the transition amplitude and $n$ refers to the final
state.  Substituting for $T_n$ and summing over final $n$ particle
states, we get
\begin{eqnarray}
\frac{\d \sigma^{s_1 s_2}}{\d x~\d Q^2} &=& \frac{\alpha^3}
{4 x s^2 Q^2}
\int_0^\infty \frac{{\d \kappa^2}}{\kappa^2}
 ~ \int_x^1 \frac{\d y}{y^2}~ L^{\mu \mu
^\prime} (q,p_1,s_1)~ L^{\nu \nu^\prime} (k,p_2,s_2)
\nonumber \\
&& \times \sum_{\lambda=0,\pm 1} g^{\lambda \lambda} ~\epsilon^*_
\nu (k,\lambda) ~\epsilon_{\nu^\prime} (k,\lambda)~ W^\gamma_{\mu
\mu^\prime} (q,k,\lambda) ~,
\end{eqnarray}
where $\alpha = e^2/4\pi$, $s$ is the centre of mass energy of
incoming leptons, $q=p_1-p_1^\prime$ and $k=p_2-p_2^\prime$
are the momenta of the probe and target photons with invariant
mass $Q^2=-q^2$ and $\kappa^2=-k^2$ respectively.  The vector
$\epsilon^\mu(k,\lambda)$ is the polarisation vector of the target
photon with polarisation $\lambda$.  The Bj\"orken variable with
respect to the target positron is defined as $x \equiv Q^2/2
\tilde \nu$, where $\tilde \nu= p_2.q$ and that with respect to
the target photon is $y \equiv Q^2/2 \nu$, where $\nu= k.q$.
The lepton tensor $L^{\mu \nu}(q,p,s_i)$ is generically defined
as
\begin{eqnarray}
L^{\mu \nu}(q,p,s_i) = 4~ p^\mu p^\nu - 2~(p^\mu q^\nu + p^\nu q^\mu)
+ 2~ p.q~ g^{\mu \nu} - 2i~ \epsilon^{\mu \nu \alpha \beta}~ s^i_\alpha
{}~q_\beta ~.
\nonumber
\end{eqnarray}
The photon structure tensor $W^\gamma_{\mu \nu} (q,k,\lambda)$ is the
imaginary part of the forward amplitude $\gamma^*(q) ~\Gamma(k,\lambda)
\rightarrow \gamma^*(q) ~\Gamma(k,\lambda)$, where $\Gamma (k,\lambda)$
is the target photon which in our analysis is treated as a non
perturbative object.  The above cross section can be written in terms
of the electron structure tensor $W^e_{\mu \mu^\prime}(q,p_2,s_2)$ as
\begin{eqnarray}
\frac{\d \sigma^{s_1 s_2}}{\d x ~ \d Q^2} =\frac{\alpha^2}{4 x^2 s^2 Q^2}~
L^{\mu \mu^\prime} (q,p_1,s_1)~ W^e_{\mu \mu^ \prime} (q,p_2,s_2) ~,
\label{TCS}
\end{eqnarray}
where $W^e_{\mu \mu^\prime}(x,Q^2,s_2)$ is
\begin{eqnarray}
W^e_{\mu \mu^\prime}(x,Q^2,s_2)=\alpha x
\int_0^\infty \frac{{\d \kappa^2}}{\kappa^2}~ \int^1_x \frac{\d y}{y^2}
L^{\nu \nu^\prime} (k,p_2,s_2) \sum_{\lambda=0,
\pm 1} g^{\lambda \lambda}~ \epsilon^*_{\nu} (k,\lambda)~\epsilon
_{\nu^\prime} (k,\lambda)~ W^\gamma_ {\mu \mu^\prime} (q,k,\lambda)~.
\nonumber
\end{eqnarray}
Due to the presence of $W_{\mu \nu}^\gamma(q,k,\lambda)$, the electron
structure tensor $W_{\mu \nu}^e(x,Q^2,s_2)$ can at best be parametrised
as spin $1/2$ target in terms of vectors $q, p_2, s_2,$ subject to the
symmetries as
\begin{eqnarray}
W^e_{\mu \nu}(x,Q^2,s_2) & =& F_1^e(x,Q^2)~ G_{\mu \nu}  + F_2^e(x,Q^2)
{}~\frac{\widetilde R_\mu\widetilde  R_\nu}{\tilde \nu}
\nonumber\\
&& + \frac{i}{\tilde \nu^2} \epsilon_{\mu \nu \alpha \beta}
\left ( g_1^e(x,Q^2)~ \tilde \nu ~q^\alpha s_2^\beta + g_2^e(x,Q^2)~
q^\alpha (\tilde \nu~ s_2^\beta - s_2 \cdot q~ p_2^\beta)\right ) ~,
\label{EST2}
\end{eqnarray}
where the tensor coefficients are defined as
\begin{eqnarray}
G_{\mu \nu} = -g_{\mu \nu} +\frac{q_\mu q_\nu}{q^2}~ , \qquad
 \widetilde R^\mu = p_2^\mu -\frac{\tilde \nu}{q^2} ~q^\mu ~.
\nonumber
\end{eqnarray}
The photon structure tensor is defined as the Fourier transform
of the commutator of $em$ currents $J_\mu(\xi)$ sandwiched between
target photon states, as
\begin{eqnarray}
W_{\mu \nu}^\gamma(k,q,\lambda) &=& \frac{1}{2 \pi} \int \d^4 \xi ~
e^{- i q\cdot \xi} ~\langle \Gamma(k, \epsilon^*(\lambda)) \vert ~
\left [J_{\mu}(\xi), J_{\nu}(0) \right ]~ \vert \Gamma(k,\epsilon
(\lambda)) \rangle_c ~,
\label{FTD}
\end{eqnarray}
where the subscript $c$ denotes the connected part.
This can be parametrised in a gauge invariant way in terms of the various
structure functions using the general symmetry arguments {\it viz.}, time
reversal invariance, parity, hermiticity and current conservation as
\begin{eqnarray}
W^\gamma_{\mu \nu} (y,Q^2,\kappa^2) &=& -\frac{1}{\kappa^4}~ \left
\{ F^\gamma_1 (y,Q^2,\kappa^2) ~G_{\mu \nu} + F^\gamma_2 (y,Q^2,
\kappa^2) ~\frac{ R_\mu  R_\nu}{\nu} \right .
\nonumber \\
&& +~ b^\gamma_1(y,Q^2,\kappa^2) ~ r_{\mu \nu}
+ ~b^\gamma_2(y,Q^2,\kappa^2) ~ s_{\mu \nu}
\nonumber \\
&& +~ b^\gamma_3(y,Q^2, \kappa^2) ~ t_{\mu \nu}  +~ b^\gamma_4(y,Q^2,
\kappa^2) ~ u_{\mu \nu}
\nonumber\\
&&\left . +~ \frac{i}{\nu^2}~ \epsilon_{\mu \nu \lambda \rho}~
\left ( g^\gamma_1(y,Q^2,\kappa^2) \nu ~q^\lambda~  s^\rho
+~ g^\gamma_2(y,Q^2,\kappa^2) ~q^\lambda~ ( \nu ~ s^\rho - s
\cdot q ~k^\rho) \right ) \right \}~,
\label{PST}
\end{eqnarray}
where $s_\mu$ is the spin vector of the target photon, $R_\mu=(x/y)
{}~\widetilde R_\mu$ and various tensors are defined as
\begin{eqnarray}
 r_{\mu \nu} &=&  \left (\frac{k\cdot E^* ~k\cdot E}{\kappa^4}
+  ~\bar \alpha^2 \right )~ G_{\mu \nu} ~, \qquad
 s_{\mu \nu} = \frac{1}{\nu} ~\left (\frac{k\cdot E^* ~k\cdot E}{\kappa^4}
+  ~\bar \alpha^2 \right ) ~ R_\mu ~ R_\nu  ~,\nonumber\\
 t_{\mu \nu} &=&\!\!\! -\frac{1}{2 \nu} \left (\frac{k\cdot E^*} {\kappa^2}
{}~(R_\mu ~E_\nu
+ R_\nu ~E_\mu) + \frac{k\cdot E}{\kappa^2} ~(R_\mu ~E_\nu^*
+ R_\nu ~E_\mu^*) + 4~(1-\bar \alpha^2) R_\mu R_\nu \right ),
\nonumber\\
u_{\mu \nu} &=& \frac{1}{\nu} \left (E^*_\mu ~E_\nu + E^*_\nu ~E_\mu
- 2~\kappa^2 ~G_{\mu \nu} - 2~(1-\bar \alpha^2) R_\mu ~R_\nu
\right ) ~,
\nonumber\\
E_\mu &=& \epsilon_\mu - \frac{q \cdot \epsilon}{\nu} ~k_\mu ~,
\qquad \bar \alpha^2 = 1- \frac{\kappa^2 Q^2}{\nu^2} ~, \qquad
s^\mu \equiv \frac{i}{\kappa^2}~\epsilon^{\mu \nu \alpha \beta} ~\epsilon^*_
\nu ~\epsilon_\alpha ~k_\beta~.
\nonumber
\end{eqnarray}
Note that the photon tensor is manifestly gauge invariant.  The additional
four structure functions $b_{1-4} ~(y,Q^2,\kappa^2)$ are due to the
off-shell nature ($\lambda =0$, scalar polarisation ($k^2 <0$)) of
the target photon.  The tensor coefficients of these additional structure
functions vanish when the photon polarisation is summed and survive when
the target is polarised and the probe polarisation is summed and hence
are called the singly polarised structure functions.  This singly
polarised nature is characteristic property of a spin one target.
In the context of photon structure function which are realised in a
$e^+ ~e^- \rightarrow e^+ ~e^- X$ process, the singly polarised part
does not manifest itself as in a spin target, but turns out to be
a part of the unpolarised cross section.  If the target photons are
real $\left (\epsilon (\lambda= \pm1) \right )$, the new singly
polarised structure functions would have been absent.

Using the above parametrisation, the unpolarised cross section for
$e^+ ~e^- \rightarrow e^+ ~e^- X$ is found to be
\begin{eqnarray}
\frac{\d \sigma^{\uparrow \uparrow +\uparrow \downarrow}}{\d x~\d Q^2}
&=& \frac{\alpha^3}{x~s^2~Q^2}~ L^{\mu \mu^\prime}_{sym} (q,p_1,\uparrow)
\int_0^\infty \frac{{\d \kappa^2}}{\kappa^2}~ \int_x^1 \frac{\d y}{y^2}~
{}~\kappa^4
\left\{ \frac{y}{x}~ {\overline P}_{\gamma e}\left (\frac {x}{y}
\right ) \sum_{\lambda=0,\pm 1} g^{\lambda \lambda}~ W^\gamma_{\mu
\mu^\prime} (\lambda) \right .
\nonumber \\
&&\left . + ~\frac{y}{x} ~\delta  P_{\gamma e} \left (\frac {x}{y}
\right ) \sum_{\lambda=0,\pm 1} C(\lambda) ~W^\gamma_{\mu \mu^\prime}
(\lambda) \right\} ~,
\label{UPCS}
\end{eqnarray}
where $C(\lambda) = 2$ for $\lambda=0$; $-1$ for $\lambda=\pm 1$.  In
the above equation, the first term in the curly bracket corresponds
to unpolarised structure function and the second term to the singly
polarised structure functions.  The modified splitting functions
($k^2 \not= 0$) are given by
\begin{eqnarray}
\overline P_{\gamma e}\left (\frac{x}{y} \right ) =\frac{y}{x}\left
(2-2 \frac{x}{y} + \frac{x^2}{y^2}\right)-2 ~\frac{y}{x}\left(1-\frac
{x}{y}\right) ~,
\nonumber\\
\delta P_{\gamma e} \left (\frac{x}{y} \right ) =\frac{y}{2 x}
\left(2-2 \frac{x}{y}+ \frac{x^2}{y^2}\right)-2 ~\frac{y}{x}\left
(1-\frac{x}{y}\right)  ~.
\nonumber
\end{eqnarray}
The first term is the usual Altarelli-Parisi splitting function
arising from the splitting of $e^+$ into transverse photons ($k^2
= 0$).  The additional term arises from the emission of zero
polarised photon. The polarised cross section is given by
\begin{eqnarray}
\frac{\d \sigma^{\uparrow \uparrow -\uparrow \downarrow}}{\d x~\d Q^2}
&=& -\frac {\alpha^3}{x~s^2~Q^2}~ L^{\mu \mu^\prime}_{asym} (q,p_1,\uparrow)
\int_0^\infty \frac{{\d \kappa^2}}{\kappa^2}~ \int_x^1\frac{\d y}{y^2}~
\kappa^4 ~
\frac{y}{x}~ \Delta P_{\gamma e}\left(\frac{x}{y} \right)
\nonumber\\
&&
\times ~\sum_{\lambda=0\pm 1} C^\prime (\lambda) W^\gamma_{\mu \mu^\prime}
(y,Q^2,\kappa^2,\lambda) ~,
\label{PCS}
\end{eqnarray}
where $C^\prime (\lambda)= 0$ for $\lambda =0$; $\pm 1$ for $\lambda=
\pm 1$ and the polarised splitting function is
\begin{eqnarray}
\Delta P_{\gamma e}\left(\frac{x}{y}\right) =\left(2-
\frac{x}{y}\right) ~.
\nonumber
\end{eqnarray}
The additional zero polarisation of the photon does not
affect the polarised cross section.

Now the electron structure functions can be related to the photon
structure functions by substituting the photon structure tensor
eqn.~(\ref{PST}) in eqn.~(\ref{UPCS},\ref{PCS}) and comparing it with the
total cross section eqn. (\ref {TCS}) after substituting eqn. (\ref
{EST2}).  Hence we get
\begin{eqnarray}
F_1^e(x,Q^2) &=& 2 \alpha
\int_0^\infty \frac{\d \kappa^2} {\kappa^2}~ \int_x^1 \frac{\d y}{y}~
\left[ \overline P_{\gamma e}\left(\frac{x}{y}\right)~
F_1^\gamma(y,Q^2, \kappa^2) \right.
\nonumber\\
&&\left.+ 2~ \delta P_{\gamma e} \left(\frac{x}{y} \right)~ \left(
\bar \alpha^2~ b_1^\gamma(y,Q^2,\kappa^2) - \frac{\kappa^2}{\nu}~
b_4^\gamma (y,Q^2,\kappa^2)\right)
\right]~, \label{F1}\\
F_2^e(x,Q^2) &=& 2 \alpha
\int_0^\infty \frac{\d \kappa^2} {\kappa^2}~ \int_x^1 \frac{\d y}{y}~
\frac{x}{y} \left[ \overline P_{\gamma e}\left(\frac{x}{y}
\right)~ F_2^\gamma(y,Q^2,\kappa^2)
\right.\label{F2}
\nonumber\\
&& \left. + 2~\delta P_{\gamma e} \left(\frac
{x}{y} \right) \left(\bar \alpha^2~ b_2^\gamma(y,Q^2,\kappa^2)
+\frac{1-\bar \alpha^2}{\bar \alpha^2} ( 1-2\bar \alpha^2)~
b_4^\gamma(y,Q^2,\kappa^2)\right)\right]~, \\
g_1^e (x,Q^2) &=& 4 \alpha \int_0^\infty \frac{\d \kappa^2} {\kappa^2}~
\int_x^1 \frac{\d y}{y}~
\left[ \Delta P_{\gamma e}\left( \frac{x}{y}\right)~ g_1^\gamma
(y,Q^2, \kappa^2) \right ]~.\label{G1}
\end{eqnarray}
Note that the unpolarised electron structure functions are related to the
singly polarised structure functions also.  This extra contribution comes
from the zero polarisation (massive nature) of the photon which is also
reflected in the modified splitting functions.  To leading order, in the
limit $\widetilde \nu \rightarrow \infty$ the unpolarised electron
structure functions $F_1^e (x,Q^2)$ and $F_2^e (x,Q^2)$ are modified
by the single polarised structure functions $b_1^\gamma (y,Q^2,\kappa^2)$
and $b_2^\gamma (y,Q^2,\kappa^2)$ respectively.  The contribution from
other singly polarised structure functions are suppressed in this limit.
The polarised structure function $g_1^e (x,Q^2)$ on the other hand is
uneffected.  We evaluate these photon structure functions to various order
in the coupling constant using the factorisation method.

In terms of the above relations we can compute the unpolarised and
polarised cross sections and are given by
\begin{eqnarray}
\frac{\d \sigma^{\uparrow \uparrow +\uparrow \downarrow}}{\d x~\d Q^2}&=&
\frac{\alpha^2}{x^2~ s~ Q^2} \left\{F_1^e(x,Q^2) ~\frac {Q^2}{s} - F^e
_2(x,Q^2) \left (1 - \frac{xs}{Q^2} \right ) \right\}~,\\
\frac{\d \sigma^{\uparrow \uparrow -\uparrow \downarrow}}{\d x~\d Q^2}&=&
2\frac {\alpha^2}{x~ s~ Q^2} ~g_1^e(x,Q^2) \left(1 -\frac{Q^2}{2xs}
\right)~.
\end{eqnarray}
Now by substituting for the electron structure functions from
eqn.~(\ref{F1}-\ref{G1}) the $n^{\rm th}$ moment of the above differential
cross section can be related to the $(n-1)^{\rm th}$ moment of the
photon structure functions.
Our next task is to understand these new structure functions in terms
of the parton distributions.

\section{Free field analysis}
One can understand the hadronic structure of these structure functions by
using the free field analysis.  In this analysis one assumes that the $em$
current $J_\mu$ to be made of free quark currents.  The leading
contribution to the commutator in eqn.~(\ref {FTD}) comes from the light
cone region $\xi^2 \rightarrow 0$.  Noting that the commutator is proportional
to the imaginary part of the time ordered product of currents and using
the Wick's expansion, we get
\begin{eqnarray}
[J_\mu (\xi),J_\nu(0)] &=& \frac{\delta^{(1)}~ (\xi^2)~ \xi^\lambda}{\pi}
\left \{ \sigma_{\mu \lambda \nu \rho}~ O^\rho_{(-)} (\xi) - i \epsilon_
{\mu \lambda \nu \rho}~ O^\rho_{(+)5} (\xi) \right \} ~,
\label{CC}
\end{eqnarray}
where
\begin{eqnarray}
\delta^{(1)}(\xi^2) & = & \frac{\partial}{\partial \xi^2}
\delta(\xi^2) ~, \nonumber\\
\sigma_{\mu \lambda \nu \rho} &=& g_{\mu \lambda}~ g_{\nu \rho} -
g_{\mu \nu}~ g_{\lambda \rho} + g_{\mu \rho}~ g_{\lambda \nu} ~,\nonumber\\
O_{(\pm)}^\rho (\xi) &=& :\overline \psi (\xi)~ \gamma^\rho~ \psi (0)~ \pm~
\overline \psi (0)~ \gamma^\rho~ \psi (\xi): ~,\nonumber\\
O_{(\pm)5}^{\rho}(\xi) &=& :\overline \psi (\xi)~ \gamma^\rho~ \gamma_5~
\psi (0)~ \pm~ \overline \psi (0) ~\gamma^\rho ~\gamma_5~ \psi (\xi):
{}~,\nonumber
\end{eqnarray}
where the : : implies normal ordering of operators.  Substituting
this in eqn.~(\ref{FTD}), performing the
$\d^4 \xi$ integral and comparing the tensor coefficients with
eqn.~(\ref{PST}), we relate the structure functions to various scaling
functions as given in Table 1.  $\widetilde A (y)$, $\widetilde B
(y)$ and $\widetilde C (y)$ in the Table 1 are defined as
\begin{eqnarray}
\int \d z^\prime ~ e^{iz^\prime k\cdot\xi} \pmatrix {\widetilde A (z^\prime )
\cr \widetilde B (z^\prime )
\cr \widetilde C (z^\prime ) } = \pmatrix {A (k\cdot\xi) \cr B (k\cdot\xi) \cr
C (k\cdot\xi) } =
\sum_n \frac{1}{(n+1)!}~  k\cdot\xi^{n-1} \pmatrix{ (n+1) A_n ~ k\cdot\xi \cr
B_n
\cr C_n ~ k\cdot\xi} ~,
\nonumber
\end{eqnarray}
where $A_n$~, $B_n$~, $C_n$ are the expansion coefficients
of local photon matrix elements given below
\begin{eqnarray}
\langle k, \epsilon^* \vert O_{(-)}^{\rho \mu_1 \cdots \mu_n} (0)
\vert k, \epsilon  \rangle &=& 2 A_n {\cal S}~ (k^\rho k^{\mu_1}
\cdots k^{\mu_n}) + B_n {\cal S} \left [ \left ({\epsilon^\rho}^*
\epsilon^{\mu_1} + {k^\rho k^{\mu_1}} \right ) k^{\mu_2} \cdots
k^{\mu_n} \right ],\\
\langle k, \epsilon^* \vert O_{(+)5}^{\rho \mu_1 \cdots \mu_n} (0)
\vert k, \epsilon  \rangle &=& C_n {\cal S} (s^\rho k^{\mu_1} \cdots
k^{\mu_n}) ~.
\end{eqnarray}
Here $A_n, B_n, C_n$ are functions of lorentz invariants such as
$k^2, s^2$ etc.  ${\cal S}$ denotes symmetrisation with respect
to all indices.  This is done to ensure only leading twist operators
contribute.  The matrix element $A_n$ contributes to the unpolarised
part, $B_n$ to singly polarised part and $C_n$ to the polarised part.
 From the Table 1 it is clear that both the unpolarised structure
functions $(F^\gamma_{1,2}~ (y))$ and the singly polarised structure
functions $(b^\gamma_{1,2}~ (y))$ satisfy Callan-Gross relation.  The
polarised structure functions $(g^\gamma_{1,2}~ (y))$ obey the Wandzura-Wilczek
sum rule.  In addition we find that $b^\gamma_3(y)$ and $b^\gamma_4(y)$ are
related
to $b^\gamma_2 (y)$ by the following relations:
\begin{eqnarray}
b^\gamma_4(y) &=& - \int_y^1 \d y^\prime ~\frac{b^\gamma_3(y^\prime)}{y^\prime}
{}~,
\nonumber \\
b^\gamma_3(y) &=& - \int_y^1 \d y^\prime ~\frac{b^\gamma_2(y^\prime)}{y^\prime}
{}~.
\end{eqnarray}

The physical interpretation of these structure functions can be given
using the above free field analysis.  This is done by substituting the
current commutator eqn.~(\ref{CC}) in eqn.~(\ref{FTD}) and performing
only the $ \d \xi^+$ and $\d \xi_\perp$ integrals, where $\xi^\pm = (\xi
^0 \pm \xi^3)/\sqrt 2$, $\xi_\perp = (\xi^1,\xi^2)$.  The unpolarised
structure functions can be extracted using the
combination $\overline W_{\mu \nu} = \sum_{\lambda=0,\pm 1} g^{\lambda
\lambda} W_{\mu \nu} $ $(\epsilon(\lambda))$, polarised by $\Delta W_{\mu
\nu} =  \sum_{\lambda= 0, \pm 1} C^\prime (\lambda) W_{\mu \nu} (\epsilon
(\lambda))$ and the singly polarised by $\delta W_{\mu \nu} =$ $ \sum_
{\lambda=0, \pm 1} C(\lambda)  W_{\mu \nu} (\epsilon (\lambda))$.
Note that the polarised
and unpolarised combinations are the usual ones while the singly
polarised combination is so chosen as to eliminate the unpolarised
and polarised structure functions.  Using appropriate projection operators
for the various structure functions, we get
\begin{eqnarray}
F^\gamma_2(y) &=& 2y ~\frac{1}{4\pi} \int \d \xi^- e^{-iy k^+ \xi^-}
\langle k \vert ~\overline O^{~+}_{(-)} (0,\xi^-,0_\perp) \vert k
\rangle~,\\
b^\gamma_2(y) &=& 4y ~\frac{1}{4\pi} \int \d \xi^- e^{-iy k^+ \xi^-}
\langle k , \epsilon^*\vert~ \delta O^+_{(-)} (0,\xi^-,0_\perp) \vert
k,\epsilon \rangle~,\\
g^\gamma_1(y) &=& \frac{1}{4\pi} \int \d \xi^- e^{-iy k^+ \xi^-}
\langle k , \epsilon^*\vert~ \Delta O^+_{(+)5} (0,\xi^-,0_\perp)
\vert k,\epsilon \rangle ~,
\end{eqnarray}
where the superscript $+$ denotes the light cone variable,
and the matrix elements in the above equations are defined
as
\begin{eqnarray}
\langle k \vert~ {\overline O}^{~+}_{(-)} (0,\xi^-,0_\perp) \vert k
\rangle &=& \sum_{\lambda =0,\pm 1} g^{\lambda \lambda}~
\langle k, \epsilon^* (\lambda) \vert~ O^+_{(-)} (0,\xi^-,0_\perp)
\vert k, \epsilon(\lambda) \rangle  ~,
\nonumber \\
\langle k, \epsilon^* \vert~ {\delta O}^+_{(-)} (0,\xi^-,0_\perp)
\vert k, \epsilon \rangle &=& \sum_{\lambda =0, \pm 1} C(\lambda)~
\langle k, \epsilon^* (\lambda) \vert~ O^+_{(-)} (0,\xi^-,0_\perp)
\vert k, \epsilon(\lambda) \rangle  ~,
\nonumber \\
\langle k, \epsilon^* \vert~ {\Delta O}^+_{(+)5} (0,\xi^-,0_\perp)
\vert k, \epsilon  \rangle &=& \sum_{\lambda=0, \pm 1} C^\prime (
\lambda) ~\langle k, \epsilon^* (\lambda) \vert ~ O^+_{(+)5} (0,\xi^-,
0_\perp) \vert k, \epsilon(\lambda) \rangle  ~. \nonumber
\end{eqnarray}
The above structure functions can be interpreted in terms of the probability
of finding a quark of helicity $h$ in a target photon of helicity $\lambda$
denoted by $f_{{\scriptstyle q}(h)/{\scriptstyle \Gamma}(\lambda)}$.
In terms of $f_{a(h)/\Gamma (\lambda)}$ the structure functions are of the
form
$F^\gamma_1 (y) = f_{{\scriptstyle a(1)}/{\scriptstyle \Gamma (0)}} -
f_{{\scriptstyle a(1)}/{\scriptstyle \Gamma (1)}} -
f_{{\scriptstyle a(-1)}/{\scriptstyle \Gamma (1)}}$,
$ b^\gamma_1(y) = 2 f_{{\scriptstyle a(1)}/{\scriptstyle \Gamma (0)}} -
f_{{\scriptstyle a(1)}/{\scriptstyle \Gamma (1)}} -
f_{{\scriptstyle a(-1)}/{\scriptstyle \Gamma (1)}}$ and
$g^\gamma_1(y) = f_{{\scriptstyle a(1)}/{\scriptstyle \Gamma (1)}} -
f_{{\scriptstyle a(-1)}/{\scriptstyle \Gamma (1)}}$~.

\section{QCD corrections: Factorisation method}
Higher order corrections (both $em$ and strong) are relevant to the
study of $\gamma^*~ \Gamma \rightarrow {\rm hadron}$ cross section
as the photonic corrections goes as $\ln Q^2$, while the QCD
corrections turns out to be of leading order itself.  Further
experimentally more than two jet events have been reported recently
\cite{LEP,AMY}.  To go beyond the leading order we make
use of the factorisation approach which is a generalisation of the free
field analysis.  This approach can be employed for the photon targets
also as its proof does not depend on the target but depends only on the
underlying theory.  This ensures
a systematic separation of hard and soft parts, {\it i.e.}
\begin{eqnarray}
\!\!\!\!\!\!W_{\mu \nu}^{\gamma^* \Gamma (\epsilon)} (y,Q^2,\kappa^2)
\!\!&=&\!\!
\sum_{a,h} \int_y^1 \frac{\d z}{z} f_{\scriptstyle a(\scriptstyle h)}
/{\scriptstyle \Gamma(\scriptstyle \epsilon)} (z,\mu_R^2,\kappa^2)~
H^{\mu \nu}_{a(h) \gamma^*} \left (q,zp,\mu_R^2,\alpha_s (\mu_R^2),
\alpha \right ) + \cdots ~,
\label{FT}
\end{eqnarray}
where the hard part $H$ of the processes are perturbatively calculable
and sum over $a$ includes partons (quarks and gluons) and free
photons. The `soft' parts are defined as photon matrix
elements of bilocal quark, gluon and photon operators as
\begin{eqnarray}
f_{{\scriptstyle q(\uparrow \downarrow)}/{\scriptstyle \Gamma(\scriptstyle
\epsilon)}}(z,\mu^2,\kappa^2) &=&
\frac{1}{4\pi} \int \d \xi^- e^{-i z ~\xi^- k^+} \langle k, \epsilon^* \vert
\overline \psi_a (0,\xi^-,0_\perp)~ \gamma^+ \Lambda_\pm
 ~{\cal G}^a_b ~ \psi^b (0) \vert k, \epsilon \rangle_c ~,
\label{Q}\\
\!\!\! f_{{\bar {\scriptstyle q}(\uparrow \downarrow)}/{\scriptstyle
\Gamma(\scriptstyle \epsilon)}}(z,\mu^2,\kappa^2) &=&
\frac{1}{4\pi} \int \d \xi^- e^{-i z ~\xi^- k^+} \langle k, \epsilon^* \vert
\overline \psi_a (0) ~\gamma^+ \Lambda_\mp
{}~ {{\cal G}^a_b}^\dagger  ~ \psi^b (0,\xi^-,0_\perp) \vert k,\epsilon
\rangle_c ~,
\label{AQ}
\end{eqnarray}
where $\Lambda_\pm = (1 \pm \gamma_5)/2$.
For $\epsilon(0)$ the $\gamma_5$ term would not be present in the
above definitions.
\begin{eqnarray}
f_{{\scriptstyle g}/{\scriptstyle \Gamma(\scriptstyle \epsilon)}}
(z,\mu^2,\kappa^2) &=& \frac{i}{4\pi z k^+} \int \d \xi^- e^{-i z ~\xi^- k^+}
\left
[\langle k, \epsilon^* \vert F_a^{+\mu} (0,\xi^-,0_\perp)~  {{\cal G}^a_b}
{}~ F^{+ b}_{\mu} (0) \vert k,\epsilon \rangle_c \right. \nonumber \\
&&\left.  + \langle k, \epsilon^* \vert F_a^{+\mu} (0)~ {{\cal G}^a_b}^\dagger
{}~F^{+ b}_{\mu} (0,\xi^-,0_\perp) \vert k,\epsilon \rangle_c \right ]~,
\label{G}\\
f_{{\Delta \scriptstyle g}/{\scriptstyle \Gamma(\scriptstyle \epsilon)}}
(z,\mu^2,\kappa^2) &=& \frac{i}{4\pi z k^+} \int \d \xi^- e^{-i z ~\xi^- k^+}
\left
[ \langle k, \epsilon^* \vert F_a^{+\mu} (0,\xi^-,0_\perp)~ {\cal G}^a_b ~
\widetilde F^{+ b}_{\mu} (0) \vert k,\epsilon \rangle_c \right . \nonumber \\
&& \left .  - ~\langle k, \epsilon^* \vert F_a^{+\mu} (0)~{{\cal
G}^a_b}^\dagger
{}~\widetilde F^{+ b}_{\mu} (0,\xi^-,0_\perp) \vert k, \epsilon \rangle_c
\right ] ~,
\label{PG}
\end{eqnarray}
where ${\cal G}^a_b = {\cal P}~\exp \left [ig \int_0^{\xi^-} \d\zeta^-
A^+(0, \zeta^-,0_\perp)\right]^a_b$, ${\cal P}$ denotes the path ordering of
gauge fields.  The above definitions hold for the photon case also where
in the $SU (3)_c$ group indices will be absent.  The gauge invariant
definitions of quark, gluon and photon distributions defined above have a
probabilistic interpretations of finding a parton or photon inside the
target photon.
These matrix elements are in principle calculable but tedious, hence we
dump all the higher order effects inside the matrix elements and treat
these as theoretical inputs.  This fact that they are calculable order
by order is exploited in the evaluation of Hard Scattering Coefficients
($HSC$).

The $\gamma^*~ \Gamma \rightarrow X$ in different energy ranges can be
classified as two and three jet events.  At low $Q^2$ (real photon),
two jet events can be explained by the VDM as well as by two photon
point interaction via quark loop (Fig.~2).  For the real photon
replacing $\Gamma = \gamma$ in the factorisation eqn.~(\ref{FT}),
the two jet point event is recovered.  When the target photon is off
mass shell and the energies are also increased, an additional two jet
event (due to the hadronic structure of photon) as shown in Fig.~3a
also contributes.  This event can be distinguished from the previous
one, as one of the jets is a spectator jet.  This diagram can also get
$em$ corrections from the bremstrahlung diagrams, which would alter the
$P_T$ of the processes.  On the other hand a gluonic correction (Fig.~3b)
gives rise to a three jet event.  Additional three jets events due to
initial state photon and gluon (Fig.~5) also exists.
We calculate the various contributions coming from these
processes to the $\gamma^* ~ \Gamma$ scattering using the above
mentioned factorisation method.

The $HSC$s can be evaluated order by order using
the factorisation formulae by replacing target photon by parton targets
{\it viz}. quarks, gluons and real photons.  We calculate the $HSC$s up
to ${\cal O}$ $(\alpha^2)$ and ${\cal O}$ $(\alpha \alpha_s)$.  Let us first
concentrate on the quark sector.  The quark sector gets contribution to
${\cal O}$ $(\alpha)$ by $\gamma^* (q)~q(p) ~\rightarrow q(p^\prime)$
(Fig.~3a), ${\cal O}$ $(\alpha^2)$ by $\gamma^* (q)~q (p)~\rightarrow
q(k) ~\gamma(k^\prime)$ and ${\cal O}$ $(\alpha \alpha_s)$ by $\gamma^*
(q)~q(p) ~\rightarrow q(k) ~g(k^\prime)$ (Fig.~3b).  For the photonic
corrections, we replace gluon lines by photon lines in Fig.~3b.  From
the factorisation formulae it is
clear that the calculation of hard scattering coefficients involves the
cross sections of the above mentioned processes as well as matrix
elements given in eqns. (\ref{Q},\ref{AQ}), with target photon replaced
by  quarks to appropriate orders.  The contributions to various structure
functions are extracted using the appropriate projection operators $P^i_{\mu
\nu}$.
Defining $W^i = P^i_{\mu \nu} W^{\mu \nu}_{\gamma^* q}$ and evaluating the
bremstrahlung diagrams given in Fig.~3b, at large $Q^2$, we get
\begin{eqnarray}
W^{F_1}_q(z,Q^2) &=& 2  f_c~ \alpha ~\left \{ -\left (\frac{1+z^2}
{1-z} \right )_+ \ln \beta_g - 2 ~\frac{1+z^2}{1-z} ~\ln z + (1+z^2) \left (
\frac{\ln (1-z)}{1-z} \right)_+ \right .\nonumber \\
&& \left . + 2z + 1 -\frac{3}{2} \frac{1}{(1-z)_+} -
\delta (1-z) \left( \frac{9}{4} + \frac{2 \pi^2}{3} \right ) \right \}
{}~,
\nonumber\\
W_q^{F_2}(z,Q^2) &=&  2z \left \{ W_q^{F_1} (z,Q^2) + 2 \alpha ~
f_c ~ 2z \right \}~,
\nonumber\\
W_q^{b_1} (z,Q^2)&=& 2 ~W_q^{F_1} (z,Q^2)  ~,\label{CSB1}\\
W_q^{b_2}(z,Q^2) &=& 2 ~ W_q^{F_2}(z,Q^2)  ~, \nonumber\\
W_q^{g_1}(z,Q^2)&=&  W_q^{F_1} (z,Q^2) - 2 \alpha ~f_c  ~ (1-z) ~,
\nonumber
\end{eqnarray}
where $z=Q^2/2p \cdot q$~, $\beta_g=m_g^2/Q^2$, $f_c =(4 \alpha_s/3, \alpha)$
is the coupling
factor depending on gluon or photon bremstrahlung respectively and the
subscript + denotes the `+ function' regularisation of the singularity as
$z \rightarrow 1$.  For the photons $m_g$ will be replaced by
$m_\gamma$.
To avoid the mass singularity, we have kept the gauge bosons massive.  If
quark masses are also kept non zero, then there would be both logarithmic and
power
singularities as these masses go to zero simultaneously.  If one of the
prescriptions is chosen, say massive gauge boson prescription, the
singularities
boil down to logarithmic singularities with an additional constant part which
depends on the prescription.  In addition there are some mass singularities
coming from the virtual diagrams which are exactly canceled by
those which arise from
regulating the bremstrahlung diagram in the limit $z \rightarrow 1$ (`+
 function').  We have considered the massive vector boson prescription
which has not so far been considered in the literature while calculating
the corrections to structure functions in general.  It has been customary
to consider the massive quark prescription or the dimensional regularisation
method to deal with the infrared (IR) singularities.

For the structure functions $F^\gamma_{1,2}(y,Q^2,\kappa^2)$ and
$b^\gamma _{1,2}(y,Q^2,\kappa^2)$ the relevant matrix elements
are obtained by replacing $\Gamma$ in eqns.~(\ref{Q},\ref{AQ}) by
quarks, where the vector part would contribute.  In the case of
polarised structure function $g^\gamma_1(y,Q^2,\kappa)$
the axial vector matrix element will contribute (eqns. (\ref{Q},
\ref{AQ})).  The contributing matrix elements are shown in Fig.~4.
The Feynman rules for the eikonal lines and vertices are given in
Ref. \cite{CSS}.  We regulate the ultraviolet (UV) divergences
appearing in these diagrams using dimensional regularisation and
keep gauge boson masses non zero to regulate the mass singularities.
Here too there is a similar cancellation of mass singularities among
the virtual and real diagrams, leaving a logarithmic singularity and
a prescription dependent constant as given bellow
\begin{eqnarray}
\!\!\!\!\!\!\!f^{(1)}_{{\scriptstyle q}/{\scriptstyle q}} (z,\mu_R^2)\!&=&\!
\frac{f_c}
{4\pi}\left \{ \left ( \frac{1+z^2}{1-z} \right )_+ \ln \beta^\prime +
\frac{1+z^2}{1-z} \ln z + 2(1-z)- \delta(1-z) \left ( \frac {9}{4} -
\frac{\pi^2}{3} \right ) \right \} ,\nonumber \\
f^{(1)}_{{\Delta \scriptstyle q}/{\scriptstyle q(\scriptstyle h)}}
(z,\mu_R^2) \!\!\! &=& \!\!\! h~ f^{(1)}_{{\scriptstyle q}/{\scriptstyle
q}}(z,\mu_R^2) ~,
\end{eqnarray}
where $\beta_g^\prime = {m_g^2}/{\mu^2_R}$ and $\mu_R$ is renormalisation
scale. The superscript $(1)$ in the above equations denotes that they are
evaluated to order $\alpha$ or $\alpha_s$ as the case may be.
This equivalence among the polarised and unpolarised matrix elements
does not hold if we keep the quark masses also non zero.
Substituting the above matrix elements and cross sections eqn.~(\ref{CSB1})
in the factorisation formulae for quark sector, we obtain
\begin{eqnarray}
H^{F_1}_q(z,Q^2) &=&  2 \alpha ~f_c~\left \{ \left (\frac{1+z^2}
{1-z} \right )_+ \ln \frac{Q^2}{\mu_R^2} -~\frac{1+z^2}{1-z} ~\ln z +
(1+z^2) \left ( \frac{\ln (1-z)}{1-z} \right)_+ \right .\nonumber \\
&& \left . + 3 -\frac{3}{2} \frac{1}{(1-z)_+} -
\delta (1-z) \left( \frac{9}{2} + \frac{\pi^2}{3} \right ) \right \}
{}~,\nonumber \\
H^{F_2}_q(z,Q^2) &=& 2z \left[H^{F_1}_q(z,Q^2) +  2 f_c ~\alpha~
2z \right] ~,\nonumber \\
H^{b_1}_q(z,Q^2) & =& 2 H^{F_1}_q(z,Q^2) ~,\\
H^{b_2}_q(z,Q^2) & =& 2 H^{F_2}_q(z,Q^2) ~,\nonumber \\
H^{g_1}_q(z,Q^2) & =& H^{F_1}_q(z,Q^2) +  2 \alpha ~f_c
(z-1)~\nonumber .
\end{eqnarray}
Note that the mass term in the logarithmic and the prescription dependent
constant term cancel among the cross section $W_q~ (z,Q^2)$ and the matrix
element $f_{\scriptstyle a/\scriptstyle b}~ (z,Q^2)$, leaving the $HSC$
independent of gauge boson mass.

Next we will discuss the three jet event coming from the initial state
photons and gluons.  The contributing subprocesses are $\gamma^* (q)
{}~\gamma (p) \rightarrow q(k) ~\bar q(k^\prime)$ and  $\gamma
^*(q)~ g(p) \rightarrow q(k)~ \bar q(k^\prime)$ (Fig.~5).  To calculate
the gluonic and photonic $HSC$s  we need the matrix element (eqns.~(\ref
{G},\ref{PG})) between photon and gluon states respectively, in addition
to the  above cross sections.  The contributions to various structure
functions from the cross section, at large $Q^2$ are
\begin{eqnarray}
W_g^{F_1} (z,Q^2) &=& 8~\alpha~ f_c \left \{ (2z^2-2z+1)
\left (\ln \frac{{\cal M}_g^2}{Q^2} + \ln \frac{z}{1-z} \right )
\right . \nonumber \\
&& \left . - \frac{p^2} {{\cal M}_g^2}z(1-z)
- \frac{2m^2 }{{\cal M}_g^2}z(1-z) + 2z^2 -2z +1 \right \} ~,
\nonumber\\
W_g^{F_2} (z,Q^2) &=& 2z \left \{ W_g^{F_1} (z,Q^2) +
8~ \alpha~f_c~(4z^2 - 4z)  \right \} ~,
\nonumber \\
W_g^{b_1} (z,Q^2) &=&  2~ \left (W_g^{F_1} (z,Q^2)
- 16 ~\alpha~f_c \frac{p^2}{{\cal M} _g^2} z^2 (1-z)^2 \right)~,\\
W_g^{b_2} (z,Q^2) &=& 2~ \left (W_g^{F_2} (z,Q^2)
- 32 z~\alpha~f_c \frac{p^2}{{\cal M} _g^2} z^2 (1-z)^2 \right )~,
\nonumber \\
W_g^{g_1} (z,Q^2) &=& 4~ \alpha~ f_c  \left \{ (1-2z)
\left (\ln \frac{{\cal M}_g^2}{Q^2} + \ln \frac{z}{1-z}
\right .\right. \nonumber \\
&& \left . \left . - \frac{p^2} {{\cal M}_g^2} z(1-z) + 1 \right )
+ \frac {2m^2 }{{\cal M}_g^2} (1-z) \right \} ~,
\nonumber
\end{eqnarray}
where ${\cal M}_g^2 \equiv m^2 - p^2 z(1-z)$.
Observe that the new singly polarised structure functions $b^\gamma_
{1,2} (z,Q^2,\kappa^2)$ differs from $F^\gamma_{1,2}(z,Q^2,\kappa^2)$ by
a term  $(2p^2 z^2(1-z)^2/ {\cal M}_g^2)$.  This extra term
is due to the gauge boson of
scalar polarisation, as a result of the off shell nature of the gauge boson.
In the parton model, care should be taken in using the above result for the
subprocess cross sections because the unphysical scalar polarisation gluon
should not be considered.  In this procedure to calculate the $HSC$s, this
step is just a technique and we have no reason to avoid the scalar
polarisation.
As we go along, it will become clear that this unphysical degree of freedom
will have no effect on the $HSC$.  The relevant matrix elements are
\begin{eqnarray}
\!\!\!\!\!f^{(1)}_{{\scriptstyle q}/{\scriptstyle g}}(z,\mu_R^2)\!\!\!
&=&\!\!\!
\frac{f_c}
{4 \pi} \left [(2 z^2 -2z +1) \ln \frac{{\cal M}_g^2}{\mu_R^2}- \frac
{2 m^2 }{{\cal M}_g^2} z(1-z) -\frac {p^2} {{\cal M}_g^2} z(1-z)+ 2 z
(1-z) \right ]~, \nonumber \\
f^{(1)}_{{\delta \scriptstyle q}/{\scriptstyle g}}(z,\mu_R^2) &=&
\frac{f_c}{4 \pi} \left
[(2 z^2 -2z +1) \ln \frac{{\cal M}_g^2}{\mu_R^2}- \frac{2 m^2 }{{\cal M}
_g^2} z(1-z) - \frac {p^2} {{\cal M}_g^2} z(1-z)\right.\nonumber \\
&& \left. - 2 \frac{p^2  }{{\cal M}_g^2} z^2 (1-z)^2+ 2 z (1-z) \right ]~,\\
f^{(1)}_{{\Delta \scriptstyle q}/{\scriptstyle g (h)}}(z,\mu_R^2)\!\! &=&\!\!
h~\frac{f_c}{4 \pi} \left [(1 -2z) \ln \frac{{\cal M}_g^2}
{\mu_R^2} + \frac{p^2}{{\cal M}_g^2} z(1-z) \right ] ~,
\nonumber
\end{eqnarray}
which are evaluated from cut diagrams shown in Fig.~6.  The details of the
calculation for polarised case can be found in Ref. \cite{PR} and unpolarised
in Ref. \cite{PM}.  Substituting the cross section and matrix elements in the
factorisation formulae, we get
\begin{eqnarray}
H_g^{F_1}(z,Q^2) &=& 8 \alpha f_c \left [ (2z^2 -2z
+1) \left ( - \ln \frac{Q^2}{\mu_R^2} + \ln \frac{z}{1-z} \right ) + 4z^2 -4z +
1 \right ] ~,
\nonumber\\
H_g^{F_2}(z,Q^2) &=& 2z \left [ H_g^{F_1} (z,Q^2) + 8
\alpha f_c(4z^2 -4z) \right ] ~,
\nonumber\\
H_g^{b_1}(z,Q^2) &=& 2 H_g^{F_1}(z,Q^2) ~,\\
H_g^{b_2}(z,Q^2) &=& 2 H_g^{F_2}(z,Q^2) ~,
\nonumber\\
H_g^{\scriptstyle g_1}(z,Q^2) &=& 4\alpha f_c\left[ (2z-1) \left (\ln
\frac{Q^2} {\mu_R^2} - \ln\frac{z}{1-z} \right ) -4z+3\right] ~.
\nonumber
\end{eqnarray}
As expected the mass terms cancel among cross section and
matrix element, so also the scalar gauge boson contribution.
Substituting the calculated $HSC$s in eqn.~(\ref{FT}) one
gets the QCD corrected $W^\gamma_{\mu \nu} (y,Q^2,\kappa^2)$ for
finite virtual target photon mass.  Choose the renormalisation
scale $\mu_R^2 = Q^2$ so that the $Q^2$ dependency of the $HSC$s
will be transferred to the strong coupling constant and the
parton distribution function $f_{a/\Gamma} (z,\mu_R^2=Q^2,\kappa^2)$.
At every order in $\alpha_s (Q^2)$, the $\ln Q^2$ growth of
$f_{a/\Gamma} (Q^2)$ is compensated by the coupling constant.
The photonic operator also contributes in the same order.  But
due to the $\ln Q^2$ growth of $f_{q/\gamma} (Q^2)$, the photon
structure tensor $W^\gamma_{\mu \nu}$ grows as $\ln Q^2$ to
leading order $(\alpha_s^0)$. The first moment of the gluonic
and photonic $HSC$ of $g_1^\gamma (z,Q^2,\kappa^2)$ vanishes
and hence are not corrected by these operators.  So the first
moment of $g_1^\gamma (z,Q^2,\kappa^2)$
is proportional to only quark field operators i.e. $f_{\Delta q/
\Gamma} (Q^2,\kappa^2)$.  For the real photon this function is
calculable and is found to be zero.  So the first moment of $g_1
^e (x,Q^2)$ is zero for real target photons.  For off shell
target photons this quantity is non zero.

\section{Conclusion}
We have analysed the $\gamma^* \Gamma \rightarrow X$ subprocess
of the $e^+ e^- \rightarrow e^+ e^- X$ process at various energy
ranges, in the light of the recent LEP and TRISTAN experiments.
The virtuality of the target photon gives rise to new singly
polarised structure functions.  We found that the unpolarised
cross section is modified bye these new structure functions to
leading order.  In addition the splitting functions are altered
by the zero polarisation of the virtual target photon.  Using a
free field analysis we have studied how the zero polarisation of
the target photon alters the physical interpretation of these
structure functions.  In the process we have got relations among
various structure functions.  In addition we have also systematically
computed various QCD and QED contributions to these structure
functions using the factorisation method.  The differential cross
section is found to grow as $\ln Q^2$ while higher order QCD
corrections are found to contribute to leading order.  Interestingly,
the first moment of $g^e_1 (x,Q^2)$ is found to be proportional to
the first moment of $f_{q/\Gamma} (Q^2,\kappa^2)$ which vanishes for
real photons.

We thank G T Bodwin for clarifying some points regarding the
regularisation scheme adopted in the context of factorisation
method.  Thanks are due to R Godbole, H S Mani, M V N Murthy,
J Pasupathy and R Ramachandran for useful discussion.  VR
thanks CTS for their hospitality while part of this work was
carried out.  We acknowledge the use of symbolic manipulation
packages {\it viz.} FORM and MACSYMA.

\eject

\eject
{\Large \bf Table}
\begin{list}
{}{\setlength{\labelwidth}{20mm}}
\item Table. 1
Various structure functions related to scaling functions and their $n^{\rm
th}$ moments.
\end{list}
\vspace{1cm}

\begin{tabular}{|c||c|c|} \hline
Structure Function & Scaling Functions & Moments \\ \hline \hline
$F_1(y)$ & $\widetilde A (y)$ & $\frac{(-i)^{n-1}}{4} A_{n-1}$ \\ \hline
$F_2(y)$ & $2y \widetilde A (y)$ & $\frac{(-i)^n}{2} A_n$ \\ \hline \hline
$b_1(y)$ & $- \frac{i}{2} y \frac{\partial^2} {\partial y^2} \widetilde
B (y)$ & $\frac{(-i)^{n-1}}{4} B_{n-1}$ \\ \hline
$b_2(y)$ & $- i y^2 \frac{\partial^2} {\partial y^2} \widetilde
B (y)$ & $\frac{(-i)^n}{2} B_n$ \\ \hline
$b_3(y)$ & $ - i y \frac{\partial} {\partial y} \widetilde B (y)$ &
$-\frac{(-i)^n}{2(n+1)} B_n$ \\ \hline
$b_4(y)$ & $- i \widetilde B (y)$ & $\frac{(-i)^n}{2n(n+1)} B_n$ \\ \hline
\hline
$g_1(y)$ & $- \frac{1}{2} y \frac{\partial} {\partial y} \widetilde
C (y)$ & $\frac{(-i)^{n-1}}{4} C_{n-1}$ \\ \hline
$g_2(y)$ & $\frac{1}{2} \widetilde C (y) + \frac{1}{2} y \frac{\partial}
{\partial y} \widetilde C (y)$ & $- \frac{(-i)^{n-1}}{4} \frac{n-1}{n}
C_{n-1}$ \\ \hline
\end{tabular}\vspace{.5cm}\\
\eject

{\Large \bf Figure Captions}
\begin{list}
{}{\setlength{\labelwidth}{20mm}}
\item  [Fig 1.]
The process $e^+ e^- \rightarrow e^+ e^- X$ via the
photon-photon interaction.

\item  [Fig 2.]
The subprocess $\gamma^* \Gamma \rightarrow q \bar q$
where the target photon $\Gamma$ is real.

\item  [Fig 3.]
(a) Born diagram, (b) Next to leading order corrections to it.

\item  [Fig 4.]
The contribution to the matrix element $f_{{\scriptstyle q}/
{\scriptstyle q}}$ and $f_{{\Delta \scriptstyle q}/{\scriptstyle q}}$
up to ${\cal O} (\alpha_s)$.

\item  [Fig 5.]
The gluon(photon) produced in the target photon
interacting with the probe photon at a higher order.

\item  [Fig 6.]
The ${\cal O} (\alpha_s)$ contribution to the matrix
element $f_{{\scriptstyle q}/{\scriptstyle g}}$ and $f_{{\Delta
\scriptstyle q}/{\scriptstyle g}}$. The vertex is $\gamma^+$ for
unpolarised and $\gamma^+ \gamma_5$ for polarised.  Here only the
non vanishing diagrams are given.

\end{list}
\end{document}